# On strategies for risk management and decision making under uncertainty shared across multiple fields

Alexander Gutfraind[1,2,3]

[1] Amazon Web Services, Inc., Herndon, Virginia, USA. [2] Loyola University Chicago, Chicago, IL, USA. [3] University of Illinois at Chicago, Chicago, IL, USA. agutfraind.research@gmail.com

## Abstract

Decision theory recognizes two principal approaches to solving problems under uncertainty: probabilistic models and cognitive heuristics. However, engineers, public planners and decision-makers in other fields seem to employ solution strategies that do not fall into either field, i.e., strategies such as robust design and contingency planning. In addition, identical strategies appear in several fields and disciplines, pointing to an important shared toolkit.

The focus of this paper is to develop a systematic understanding of such strategies and develop a framework to better employ them in decision making and risk management. The paper finds more than 110 examples of such strategies and this approach to risk is termed RDOT: Risk-reducing Design and Operations Toolkit. RDOT strategies fall into six broad categories: structural, reactive, formal, adversarial, multi-stage and positive. RDOT strategies provide an efficient response even to radical uncertainty or unknown unknowns that are challenging to address with probabilistic methods. RDOT could be incorporated into decision theory using workflows, multi-objective optimization and multi-attribute utility theory.

Overall, RDOT represents an overlooked class of versatile responses to uncertainty. Because RDOT strategies do not require precise estimation or forecasting, they are particularly helpful in decision problems affected by uncertainty and for resource-constrained decision making.

**Keywords**: decision theory, radical uncertainty, probabilistic risk assessment, heuristics, multi-objective optimization, risk management

# Introduction

Resolving most decision or risk management problems requires addressing uncertainty in all its forms [1,2]. While uncertainty can be modeled and managed using probability theory [3–5], some problems are particularly difficult to address in practice. This encompasses problems where (1) It is challenging to estimate the necessary model parameters [6], (2) Uncertainty is due to ambiguity or under-determinacy [7–9]; and (3) The problem involves low-probability, high-consequence events [10,11]. Some authors go further and propose that certain decision problems are affected by uncertainty that is resistant to quantitative methods. This idea has been termed "Knightian uncertainty" [12], "Hard uncertainty" [13], "Unknown Unknowns" [14], "Black Swans" [15] and, most recently, "Radical uncertainty" [16] – all with broadly similar meaning (for responses see [10,17–19]).

How do decision-makers cope with uncertainty? An influential research program originating with Herbert Simon [20] and continued in the behavioral economics literature [21,22] investigated the feasibility of heuristics for decision making. In their words, "[heuristics] describe simple decision processes that use limited information" [23,24]. Rather than dismissing heuristics as suboptimal, they were proven to be effective and sometimes more effective than models that could be built under realistic conditions. For example, investors that split their bets evenly across a portfolio of N assets outperformed investors that estimated asset performance using limited data [25].

Expert practitioners in areas such as engineering, public policy and others also routinely face complex decisions under uncertainty. What strategies do *they* use? About 10 years ago, I interviewed experts in the two seemingly-unrelated fields of counter-terrorism and electrical engineering about their design strategies in the face of uncertainty. I found that they employ remarkably similar strategies: in both fields, the practitioners employ the strategies of multi-layer defense [26], robust design [27] and resilience [28]. These are powerful strategies in that they can succeed even against hard-to-predict events and require only readily-available data and computational resources. Yet, they are not heuristics in the economic or psychological senses above, i.e., having a clear search rule, a stopping rule, and a decision rule [23,24]. Additionally, they depart from the decision setting considered in the heuristics literature because they often involve (1) large-scale efforts of teams and organizations, rather than a single decision-maker; (2) complex decision-making settings such as system design and competitions, rather than choices from existing alternatives; and (3) sophisticated computational methods applied over weeks, months or years, rather than rules applied on the spot.

By extending from these interviews, this paper seeks to collect and describe additional strategies for decision making under uncertainty. It reports on more than one hundred universal strategies, i.e. strategies with two or more different fields of application. The paper then develops an ontology of this toolkit and incorporates it into the theory of rational decision making. The paper is most closely related to the program of Todinov [29] that sought to find generic solutions to risk, but with broader applications. Those studies were limited to risks in technical systems, while this paper includes strategies for all types of uncertainty – including decision making and organizational problems – and draws on solutions from fields such as management and military sciences. Overall, this paper aims to advance the theory of an important class of solutions to uncertainty and increase their use.

# Decision settings

This study collects strategies across diverse decision settings affected by uncertainty (as described in Methods below). It considers cases where the decision-makers are individuals, organizations or public authorities. The settings might involve decisions with a small set of well-defined alternatives, open-ended design questions, or game-theoretic multi-agent problems. The goal of the decision-maker might be system optimization, system design, improvement of planning or response, or reduction of risk.

The study included a variety of different decision scenarios. In some of the decisions, the setting can be modeled using the states-actions-outcomes framework [30]. Namely, the decision maker (or makers) face an uncertain current state of the world, $s \in S$ and have to choose between actions $a \in F$ from a set that might be infinite. An action is a function that maps states of the world to outcomes (also called objectives): $G_a(s) \in O$. Our decision-makers are often unaware of the possible actions and outcomes [31], unlike in classical settings [32]. The action set might be very small (e.g. choosing between two options) or, as happens in problems of system design, very large. The decision might occur at a single point in time, in multiple stages, or as an infinite sequence of decisions. Solving a problem usually means finding a "satisficing" solution in the tradition of Herbert Simon [20,23] and only in a few cases would the solution be optimal under an objective function.

In some of the cases I encountered below, the decision setting was analogous to the system-environment setting of stochastic control theory [33]. It consists of the following four components: (1) the system – the physical process or entity being controlled and (2) environment – external elements that affect the system's operations and success. The decision maker (or makers) have influence over (3) certain control inputs and seek to attain (4) outputs or goals. The decision maker has incomplete information about the present or future states of the environment and of the system itself. Often, only part of the system is observed and there is also uncertainty about the laws governing its evolution. Decisions are often made in the presence of other actors (e.g. peers, assistants or adversaries) whose actions are poorly predicted. The decision strategies discussed below then seek to reduce one or more of these types of uncertainty.

# Methods

## Search methodology

Because uncertainty is a vast challenge across many fields, the likely number of response strategies is large. Therefore, in this study the goal was not to attempt to build an encyclopedic list of such strategies across multiple fields but rather to find a large set of diverse examples and to extrapolate a general framework from these cases. The strategies were primarily sourced from articles, monographs and interviews. Important examples came from monographs across diverse fields: military science [34], project management [35], safety and reliability engineering [36], software DevOps and security [37] and biology [38]. I also contributed examples from personal experience, primarily in engineering, software development, consulting and counter-terrorism

In a typical workflow, reading of the literature would yield an interesting candidate decision-making technique supported by an example. It was considered relevant if it either aimed to reduce uncertainty or risk, or if it was designed to select among alternatives under uncertainty. Also included were strategies that harness positive outcomes of uncertainty (e.g. real options), as well as strategies that help in adversarial situations (e.g. misdirection).
Next, the candidate strategy would be compared to existing strategies in the collection to ensure its novelty. If novel, the strategy would be defined precisely and a search would start for examples of this strategy from a different field, ideally not closely related. An online repository was created to accumulate the strategies over time (see Data Availability Statement). All strategies were supported by references to the professional literature, with up to three examples and links to any related strategies.

After about 60 strategies were found, they were compared to each other in order to define categories of strategies. The categories went through multiple revisions to better define the boundaries between them, ultimately resulting in six categories. Subsequently, each new strategy was assigned to one or more of these categories.

# Results

Search for strategies netted a total of over 110 strategies, listed in full as Supplemental Data. The strategies are clearly defined and listed in a table alongside applications, notes and citations. Taken together, the list is a toolkit for risk management and decision making under uncertainty that I called Risk-reducing Design and Operations Toolkit (RDOT). The strategies could be organized into six categories (Table 1).

**Table 1: Ontology of 110 solution strategies for uncertainty.** Count refers to the current RDOT catalog.

| Category | Count | Subcategories |
|---|---|---|
| Formal strategies involving algorithms or workflows | 42 | (a) Algorithmic (computational) methods (b) Analytical methods |
| Structural strategies that design for uncertainty | 39 | (a) Design of technological systems (b) Design of social systems |
| Reactive strategies that improve responses to events | 10 | (a) The detection of events (b) The speed and efficiency of response |
| Strategies for adversarial situations | 11 | - |
| Strategies for multi-stage decisions | 9 | - |
| Strategies that harness positive outcomes | 8 | - |

## Formal strategies involving algorithms or workflows

Formal strategies rely on computational algorithms or systematic decision-making processes. This is the largest category of strategies, and it includes methods for risk discovery and reduction, and tools for decision and optimization in the face of uncertainty.

There are two major subcategories of formal strategies:

**Algorithmic methods** – strategies involving computation to reduce uncertainty. Some diverse examples are:

1. Mechanistic modeling - mathematical modeling that uses known physical principles to describe the behavior of a system and better understand its behavior or forecast its evolution [39]
2. Reinforcement learning - a learning strategy where an agent learns to behave in an environment by trial and error [40]
3. Portfolio rebalancing - regularly review and adjust a portfolio of items [41]

**Analytical and organizational methods** – strategies implemented as workflows for human decision-makers and organizations. Interesting diverse examples are:

1. Hypothetico-deductive method - a method that uses inductive reasoning to make predictions and then tests those predictions with experiments [42]
2. Event tree analysis - use cause-and-effect mapping to find risks by identifying possible initiating events and exploring their consequences [43]
3. Operator checkrides - require all new operators to be evaluated by an experienced co-pilot [44]

## Structural strategies: design methods for uncertainty

Structural strategies design or modify the at-risk system or organization to make it better prepared for uncertainty. There are two subcategories: technical and organizational.

**Technical** – strategies that design, modify or configure the technological system to make it more robust to hazards. These strategies are commonly used in engineering to strengthen mechanical, electrical, or other structural systems against unexpected shocks. Their goal is to prevent damage or to prevent catastrophic damage. They do not attempt to anticipate the impacting event but might incorporate knowledge about its likely type or magnitude.

Interesting examples of this subcategory include:

1. Resilient design - a system designed to absorb, respond to, and recover from disasters and adapt to new conditions [28]
2. Multi-layer defense - a system with multiple layers of barriers [26]
3. Sacrificial part - a system where damage to critical components is reduced through the sacrifice of less critical components [45]
4. Evolvable design - a design that can be easily changed and adapted over time [44,46]

**Organizational** – strategies that prepare the organization for uncertainty or hazards. Examples of such strategies include:

1. Contingency planning - establishing thorough plans, procedures, and technical measures that can enable a system to be recovered as quickly and effectively as possible following a service disruption [47]
2. Delegate (or centralize) control - improving the system by reassigning powers within the organization, e.g. mission command in the military [48]
3. Regulation by state or non-state authorities - either limiting risk-producing activities or implementing processes to reduce risks from them [49]

## Reactive strategies that enable a better, faster, or more successful response

These strategies involve improving responses to active events. The two categories are detection and post-detection strategies.

**Detection** - strategies that increase the probability and/or rate of detection including:

1. Early warning system - predict and detect possible hazardous events in advance [50]
2. Anomaly detection system - seek out anomalous readings and trigger alerts e.g. [51]

3. Event forensics and attribution - determine the type of an event and reduce uncertainty about its perpetrators and consequences, e.g. in nuclear detonations [52]

**Post-detection** - strategies that increase the speed or efficiency of response

1. Stand-off response system - interdicting impacting events at a safe distance e.g. [53]

2. Incident response unit - establish an organization dedicated to responding to damaging events e.g. [54]

3. Automatic containment system - a system that automatically contains the damage or fault propagation [55]

Comparison of reactive strategies to structural strategies, I note that both require design and planning but reactive strategies do not fundamentally change the design of the system at risk.

As a generalization, structural strategies tend to be passive and ever-present while reactive strategies involve fast and powerful activities around the time of the event. Reactive measures are found in larger, more complex or more intelligent systems where they are added to a foundation of structural measures.

In addition to these categories, there are strategies that work in three special situations: adversarial, multi-stage and positive outcomes. These are primarily formal methods, and are often used in combination with other strategies in these special situations.

## Strategies for adversarial situations

These strategies are designed to assist against uncertainty created by an adaptive adversary, such as a business competitor, political or military enemy, or an adaptive algorithm. Examples of solutions include:

1. Accelerate adaptation - learning and quickly adjusting tactics [56]

2. Misdirection - cause the adversary to take incorrect action with a false impression[57]

3. rK strategy - outcompete by rapid reproduction or focusing of resources [58]

## Strategies for multi-stage decisions

These strategies aid improving outcomes in multi-stage decisions under uncertainty, i.e. sequences of decisions spread over time. Often investment in the first stage can enable new, more favorable options in subsequent stages, including some of the strategies mentioned earlier. Examples of such strategies are:

1. Basic research - deepen understanding of the phenomenon to enable solutions
2. Index building - develop a measurement of the phenomenon to enable statistical analysis or financial products [59]
3. Prototype-driven development - build design prototypes in order to learn more about it [60]

## Strategies that harness positive outcomes

Uncertainty sometimes represents positive rather than negative outcomes, and these strategies attempt to harness them. Interesting examples from this category include:

1. Real options - purchase positive risk, specifically real options or financial instruments [61]
2. Grow-the-funnel - increase the number of trials in order to obtain more wins, e.g. developing leads for future sales [62]
3. Unicorn hunting - in selecting lotteries (i.e. bets), seek lotteries that sometimes give extreme positive returns even when the typical pay has low value [63]

# Techniques for selecting and applying RDOT strategies

Applying RDOT strategies in a concrete risk management project requires selecting one or several of the strategies and optimally allocating investments between them. If expected utility could be calculated for the different strategies, as in the classical setting of decision theory, then it is possible to solve the allocation problem by maximizing the expected utility. However, in general it may be hard to compute the expected utility (as discussed in the Introduction) and an alternative solution would be required. Fortunately, there are both practical and quantitative techniques to solve this challenge. As a practical solution, decision-makers can choose strategies based on criteria such as professional expertise, familiarity and external criteria (regulations, standards, or guidelines). Additionally, there are both workflow and computational solutions for applying RDOT.

For the practicing system designers and analysts working on a novel project, the following workflow may be used. First, evaluate the project along the following dimensions: (1) the magnitude of the risk (e.g. the cost and impact of failure), (2) the available resources for the project, (3) and whether adversarial threats are material. These dimensions can rapidly narrow down the strategies that need to be considered from the 100+ in RDOT to a handful that would suit the project. In the case of smaller projects, the most suitable mitigation strategies tend to be structural and passive. Projects of higher complexity or system value should consider additional categories of strategies: when the project is embedded in large organizations, risk management should include assisting these organizations in the safe use and risk management. It also may be cost-effective to invest in reactive risk management, and strategies such as anomaly detection and containment. When more resources are available, it may be cost-effective to implement formal strategies including hazard identification strategies and post-release monitoring. Given sufficient time it may be possible to utilize multi-stage strategies such as research and data gathering to inform future risk solutions. Lastly, when the project is threatened by adaptive actors (e.g. malicious insiders, business competitors, out-of-control reasoning systems), then counter-adversarial strategies should be considered.

## Workflows for selecting strategies

An existing formal workflow that could be utilized for selecting strategies is the Risk and Control Matrix (RACM) framework. RACM is a technique for risk management that involves building a grid of undesirable events as rows, with corresponding controls that could mitigate them [64]. The term "control" refers to any activity reducing a source of risk, and it might include monitoring, audits, technical measures and others. RACM can conveniently utilize RDOT strategies, since many RDOT strategies are risk controls. A decision-maker facing a risk management project would draw the RACM and indicate in cell $(i, j)$ if risk $i$ is controlled by strategy $j$. The approach can be modeled as a minimum set cover problem [65] to compute a combination of strategies that most efficiently controls the risks and satisfies feasibility and other constraints. The computed solution should be reviewed because there are often undocumented constraints that might make the solution infeasible.

Additionally, a computational solution using Generative AI technology (GenAI) can help finding relevant strategies. In this approach, a GenAI-powered program finds possibly relevant strategies to a given project and reports on how each component of a complex project or system could be projected from hazards. The program takes two inputs: (a) the list of strategies as a table including the definition and a few examples of each strategy (i.e. the Supplemental Data), (b) a detailed description of the project including any system components and their linkages, and a description of the hazards. A program invokes a large language model (LLM) and prompts it to find possibly relevant strategies for each component of the system. The prompt is a brief instruction such as "for each system component above, find 3-5 strategies from the list of strategies that could reduce the risk to the component".

The author experimented with applying GenAI to finding risk mitigation strategies to multiple systems from different fields and found that GenAI can generate interesting suggestions. In another experiment, GenAI was applied to build risk mitigations to a fictional system (e.g. space mining installations), in order to ensure that GenAI solutions are not based on any LLM-memorized solution. In all experiments, the GenAI recommendations were plausible and useful, particularly when drawing on powerful but less popular strategies, but also invariably superficial and infeasible. A practicing risk analyst can take some inspiration from GenAI solutions and then use her knowledge and experience to design an integrated risk management solution that would be both feasible and efficient.

## Optimizing choice of RDOT strategies with multi-objective optimization

The multi-objective optimization framework can be applied with RDOT to solve decision and risk management problems in a very general way. The key idea is that even if we cannot compute the expected outcomes due to uncertainty, we can often compute surrogate metrics of performance and consider them as separate dimensions. As a hypothetical example, suppose we want to design an uncrewed spaceship, with its sole key performance indicator being internal cargo volume. To reduce the risk from solar flares, we are considering adding shielding to the electronics. While it is difficult to measure the effect of the shielding on the risk to the mission, we can measure the shield's attenuation [66] and compare the designs based on both cargo volume and shielding.

Formally, suppose without loss of generality that our problem is focused on maximizing certain system performance indicators subject to constraints:

$$\max_{x}\{K_1(x),..,K_p(x)\} \; such \; that \; P(x) \leq Q. \tag{1}$$

where $x$ is the action and $Q$ is a vector of constants bounding the vector $P(x)$ (capital letters with parentheses indicate vector-valued functions of the decision variables, and subscripts indicate their scalar components). As noted above, it is often possible to quantify how an action $x$ would change certain metrics of interest (e.g. resilience, knowledge), even if the metrics cannot be mapped to ultimate outcomes (e.g. future yield of an investment). For example, in resilience-based approaches, it is possible to quantify the system's resilience and apply computational optimization to increase the resilience [28]. Similarly, the robustness of a system as a function of $x$ could be quantified using mechanical modeling, while detection strategies might be measured by statistical methods.

Therefore, we let the relationship between action and outcomes be the mapping $x \rightarrow \{L_1(x), .., L_q(x)\}$. To decide between investments, the decision-maker would formulate an extended multi-objective problem:

$$\max_x \{K_1(x), .., K_p(x), L_1(x), .., L_q(x)\} \; such \; that \; P'(x) \leq Q' \quad (2)$$

where $Q'$ and $P'$ include $P$ and $Q$ as well as possibly additional constraints.

There are various techniques for solving such multi-objective optimization problems. In some cases, it is possible to scalarize the objectives into a scalar utility function and solve to optimality. Techniques from multi-attribute decision theory [67], such as Analytical Hierarchical Process [68], could also be applied. Other approaches include finding the efficient solution frontiers by eliminating all dominated solutions and then heuristically selecting the tradeoffs using the knee heuristic [69].

# Discussion and Conclusions

This study is a reexamination of uncertainty in decision theory and risk. It is inspired by the paradoxical situation where expert practitioners routinely make critical decisions under uncertainty while ignoring the mandate of decision theory to calculate the expected utility of actions. Moreover, the decisions often involve novel situations involving scarce information and where forecasting is difficult. As the paper tries to show, the paradox is resolved by noting that the practitioners have developed a large toolkit of solution strategies, termed RDOT, which could be applied even in the absence of any ability to forecast outcomes. Indeed, as I previously argued, the analytical complexity of decision problems is only partly related to the amount of uncertainty in it [70].

The strategies are grouped into six categories. The largest groups are "formal" strategies that chips uncertainty away or helps find optimizing solutions. Other clusters of strategies increase a system's robustness or reaction time to the unexpected. There are also strategies for special situations like adversarial and multi-stage problems, among others. It is notable that more than 100 shared strategies are found in fields as disparate as engineering, medicine and military science, and there are likely others.

To summarize, the paper identifies a sizable toolkit of strategies to address uncertainty and make decisions under uncertainty. The toolkit is field-agnostic and could be particularly valuable for emerging fields of decision making, such as artificial intelligence risks. Emerging fields could adapt and apply strategies from the RDOT collection, rather than re-inventing them (but of course, develop and complement them with unique field-specific solutions). On a theoretical level, RDOT strategies can be incorporated into a formal framework using multi-objective optimization and its variants, or as analyst workflows. Future work should explore additional bridges between probabilistic, heuristic and RDOT-type solutions to decision making.

# Acknowledgements

I would like to thank Profs. Vicki Bier, Michael Genkin and several colleagues for the helpful discussion that inspired and refined this study, and Ms. Alisa Ungar-Saron for editing. The project was sponsored in part by NIH grant R01-AI158666. The author is an employee of Amazon Web Services (AWS). The author declares that their employer had no involvement in the design, execution, or writing of this research but has reviewed the final manuscript.

# Data Availability Statement

The datasets collected in the current study are available permanently from Zenodo at https://doi.org/10.5281/zenodo.14276879. The latest versions of the RDOT catalog are stored at https://github.com/sashagutfraind/uncertainty_strategies.